\documentclass[showpacs,aps,prd]{revtex4}
\input epsf

\begin{document}

\title{The $P$-wave $[cs][\bar{c}\bar{s}]$ tetraquark state: $Y(4260)$ or $Y(4660)$?}
\author{Jian-Rong Zhang and Ming-Qiu Huang}
\affiliation{Department of Physics, National University of Defense
Technology, Hunan 410073, China}
\date{\today}

%%%%%%%%%%%%%%%%%%%%%%%%%%%%%%%%%%%%%%%%%%%%%%%%%%%%%%%%%%%%%%%%%%%%%
\begin{abstract}
The mass of $P$-wave $cs$-scalar-diquark $\bar{c}\bar{s}$-scalar-antidiquark state
is computed in the framework of QCD
sum rules. The
result $4.69\pm0.36~\mbox{GeV}$ is in good agreement with the experimental value of
$Y(4660)$ but higher than $Y(4260)$'s, which supports the $P$-wave
$[cs][\bar{c}\bar{s}]$ configuration for $Y(4660)$ while disfavors the
interpretation of $Y(4260)$ as the $P$-wave
$[cs][\bar{c}\bar{s}]$ state.
In the same picture, the mass of $P$-wave $[bs][\bar{b}\bar{s}]$
is predicted to be $11.19\pm0.49~\mbox{GeV}$.
\end{abstract}
\maketitle

%%%%%%%%%%%%%%%%%%%%%%%%%%%%%%%%%%%%%%%%%%%%%%%%%%%%%%%%%%%%%%%%%%%%%
\section{Introduction}\label{sec1}
Fruitful heavy hadrons have been observed by far,
some of which attribute to the
$J^{PC}=1^{--}$ family, e.g. $Y(4260)$, $Y(4360)$, and $Y(4660)$.
The observation of $Y(4260)$ was first announced by BABAR Collaboration \cite{Y4260-BABAR},
which was confirmed later by both
CLEO Collaboration \cite{Y4260-CLEO} and Belle Collaboration \cite{Y4260-Belle}. A fit to the
resonance yields a
mass $4263^{+8}_{-9}~\mbox{MeV}$ \cite{PDG}. Subsequently,
$Y(4360)$ \cite{Y4325,Y4360-fit,Y4360} and $Y(4660)$ \cite{Y4360} were reported by
BaBar Collaboration and Belle Collaboration,
masses of which are $4361\pm9\pm9~\mbox{MeV}$ and $4664\pm11\pm5~\mbox{MeV}$, respectively.
Since then, these states have inspired intensive theoretical
speculations. Concretely, $Y(4260)$ is proposed as
a hybrid charmonium \cite{Y4260},
a $\chi_{c}\rho^{0}$ molecular state \cite{Y4260-Liu},
a conventional $\Psi(4S)$ \cite{Y4260-FJ},
an $\omega\chi_{c1}$ molecular state \cite{Y4260-Yuan},
a $\Lambda_{c}\bar{\Lambda}_{c}$ baryonium state \cite{Y4260-Qiao}, and
a $D_{1}D$ or $D_{0}D^{*}$ hadronic molecule \cite{Y4260-Ding};
$Y(4360)$ is interpreted as the candidate of the charmonium
hybrid or a $3^{3}D_{1}$ $c\bar{c}$ state \cite{Ding};
$Y(4660)$ is suggested to be a $5^{3}S_{1}$ charmonium \cite{Ding}, a baryonium
state \cite{Qiao,Y4660-Bugg}, a $f_{0}(980)\Psi'$ bound state \cite{Guo,zgwang},  a $6^{3}S_{1}$ state \cite{KTChao},
and a $5^{3}S_{1}-4^{3}D_{1}$
mixing state \cite{Badalian}. Besides, many other renewed works \cite{renewed}
have appeared continually.

In the tetraquark picture, $Y(4260)$ is deciphered as
the $P$-wave $[cs][\bar{c}\bar{s}]$ state \cite{Y4260-L}
(named as $Y_{[cs]}$ here), however, some authors do not
go along with the assumption and figure that $Y(4260)$ cannot be a $P$-wave
charm-strange diquark-antidiquark \cite{Y4660-Ebert}. Otherwise,
some researchers study $Y(4660)$ as a charm-strange
tetraquark state \cite{Y4660-QCDSR}. Under such a circumstance, it is interesting and necessary
to make clear whether
$Y(4260)$ can be interpreted as the $P$-wave $[cs][\bar{c}\bar{s}]$ state
or $Y(4660)$ can be a candidate of the $Y_{[cs]}$.
Indubitably, the quantitative investigation of $Y_{[cs]}$'s mass is
very instructive for comprehending its structure, but it is quite
difficult to extract hadronic spectrum information from the QCD basic theory. Fortunately, one can make use of QCD sum rules
\cite{svzsum} (for reviews see
\cite{overview,overview1,overview2,overview3} and references
therein), which is entrenched in the QCD first principle. Just in this work, we devote to reckon the mass of
$Y_{[cs]}$ through the QCD sum
rule, to study whether $Y(4260)$ or $Y(4660)$ can be a
$P$-wave $[cs][\bar{c}\bar{s}]$ state.
In addition, $Y_{b}(10890)$ \cite{Y10890,Y10890-new} has been interpreted
as a $P$-wave $[bq][\bar{b}\bar{q}]$ tetraquark
state \cite{tetraquark}. Similarly, the bottom counterpart
$[bs][\bar{b}\bar{s}]$ for $Y_{[cs]}$ could exist,
thereby $Y_{[bs]}$'s mass is also predicted here.

The paper is planned as follows. The QCD sum rule
for the tetraquark state is introduced in Sec. \ref{sec2}, and both the
phenomenological and QCD side are derived, followed
by the numerical analysis and some discussions in Sec.
\ref{sec3}. Section \ref{sec4} is a brief summary.
%%%%%%%%%%%%%%%%%%%%%%%%%%%%%%%%%%%%%%%%%%%%%%%%%%%%%%%%%%%%%%%%%%%
\section{the $P$-wave $[Qs][\bar{Q}\bar{s}]$ QCD sum rule}\label{sec2}
The QCD sum rule bridges the gap between the hadron phenomenology and the
quark-gluon interactions. By analogy with the structure of $P$-wave $[Qq][\bar{Q}\bar{q}]$ in Ref. \cite{diquark},
the $Y_{[Qs]}$ is a $J^{PC}=1^{--}$ bound diquark-antidiquark state
having the flavor content $Y_{[Qs]}=[Qs][\bar{Q}\bar{s}]$ with the spin and orbital momentum
numbers: $S_{[Qs]}=0$, $S_{[\bar{Q}\bar{s}]}=0$, $S_{[Qs][\bar{Q}\bar{s}]}=0$, and $L_{[Qs][\bar{Q}\bar{s}]}=1$.
For the interpolating current, a derivative could be included in order to generate $L_{[Qs][\bar{Q}\bar{s}]}=1$.
Presently, one constructs the tetraquark state current
from diquark-antidiquark
configuration of fields, while constructs the molecular state current from
meson-meson type of fields.
While these two types of currents can be related to each other by Fiertz rearrangements,
the relations are suppressed
by a typical color and Dirac factor so that one could
obtain a reliable sum rule only if one has chosen the appropriate current
to have a maximum overlap with the physical
state (on this point, there are some calculations and discussions
in the XII. Appendix in Ref. \cite{overview-MN}).
Concretely,
it will have a maximum
overlap for the tetraqurk state using the
diquark-antidiquark current and
the sum rule can reproduce the physical mass well,
whereas the overlap for the tetraqurk state employing
a meson-meson type of current will be
small and the sum rule will not be able to reproduce the
mass well.
Thus, the following form of current could be
constructed for $Y_{[Qs]}$,
\begin{eqnarray}
j^{\mu}=\epsilon_{abc}\epsilon_{dec}(s_{a}^{T}C\gamma_{5}Q_{b})D^{\mu}(\bar{s}_{d}\gamma_{5}C\bar{Q}_{e}^{T}).
\end{eqnarray}
Here the index $T$ means matrix
transposition, $C$ is the charge conjugation matrix, $D^{\mu}$ denotes the covariant derivative, as well as $a$, $b$,
$c$, $d$, and $e$ are color indices.

To derive the mass sum rule, one starts from the two-point correlator
\begin{eqnarray}
\Pi^{\mu\nu}(q^{2})=i\int
d^{4}x\mbox{e}^{iq.x}\langle0|T[j^{\mu}(x)j^{\nu+}(0)]|0\rangle.
\end{eqnarray}
Lorentz covariance implies that the two-point correlator
can be generally parameterized as
\begin{eqnarray}
\Pi^{\mu\nu}(q^{2})=(\frac{q^{\mu}q^{\nu}}{q^{2}}-g^{\mu\nu})\Pi^{(1)}(q^{2})+\frac{q^{\mu}q^{\nu}}{q^{2}}\Pi^{(0)}(q^{2}).
\end{eqnarray}
The part of the correlator proportional to $g_{\mu\nu}$ is
chosen to attain the sum rule here. Phenomenologically,
$\Pi^{(1)}(q^{2})$ can be expressed as
\begin{eqnarray}
\Pi^{(1)}(q^{2})=\frac{[\lambda^{(1)}]^{2}}{M_{H}^{2}-q^{2}}+\frac{1}{\pi}\int_{s_{0}}
^{\infty}ds\frac{\mbox{Im}\Pi^{(1)\mbox{phen}}(s)}{s-q^{2}}+\mbox{subtractions},
\end{eqnarray}
where $M_{H}$ denotes the mass of the hadronic resonance. In the OPE
side, $\Pi^{(1)}(q^{2})$ can be written as
\begin{eqnarray}
\Pi^{(1)}(q^{2})=\int_{(2m_{Q}+2m_{s})^{2}}^{\infty}ds\frac{\rho^{\mbox{OPE}}(s)}{s-q^{2}},
\end{eqnarray}
where the spectral density is given by
$\rho^{\mbox{OPE}}(s)=\frac{1}{\pi}\mbox{Im}\Pi^{\mbox{(1)}}(s)$.
After equating the two sides, assuming quark-hadron duality, and
making a Borel transform, the sum rule can be written as
\begin{eqnarray}\label{sumrule}
[\lambda^{(1)}]^{2}e^{-M_{H}^{2}/M^{2}}&=&\int_{(2m_{Q}+2m_{s})^{2}}^{s_{0}}ds\rho^{\mbox{OPE}}(s)e^{-s/M^{2}}.
\end{eqnarray}
Eliminating the hadronic coupling constant $\lambda^{(1)}$, one could
yield
\begin{eqnarray}\label{sum rule}
M_{H}^{2}&=&\int_{(2m_{Q}+2m_{s})^{2}}^{s_{0}}ds\rho^{\mbox{OPE}}s
e^{-s/M^{2}}/
\int_{(2m_{Q}+2m_{s})^{2}}^{s_{0}}ds\rho^{\mbox{OPE}}e^{-s/M^{2}}.
\end{eqnarray}

For the OPE calculations, one works at leading order in $\alpha_{s}$
and considers condensates up to dimension six, with
the similar techniques developed in \cite{technique,technique1}. The $s$ quark is dealt as a light one and the
diagrams are considered up to the order $m_{s}$. To keep the heavy-quark mass finite, one uses the
momentum-space expression for the heavy-quark propagator, and
the expressions with two and three gluons
attached \cite{reinders} are used. The light-quark part of the correlation function
is calculated in the
coordinate space and then Fourier-transformed to the momentum
space in $D$ dimension. The resulting light-quark part is combined
with the heavy-quark part before it is dimensionally regularized at
$D=4$. Finally with

\begin{eqnarray}
\rho^{\mbox{OPE}}(s)&=&\rho^{\mbox{pert}}(s)+\rho^{\langle\bar{s}s\rangle}(s)+\rho^{\langle\bar{s}s\rangle^{2}}(s)+\rho^{\langle
g\bar{s}\sigma\cdot G s\rangle}(s)+\rho^{\langle
g^{2}G^{2}\rangle}(s)+\rho^{\langle g^{3}G^{3}\rangle}(s),\nonumber\\
\rho^{\mbox{pert}}(s)&=&-\frac{1}{3\cdot5\cdot2^{11}\pi^{6}}\int_{\alpha_{min}}^{\alpha_{max}}\frac{d\alpha}{\alpha^{4}}\int_{\beta_{min}}^{1-\alpha}\frac{d\beta}{\beta^{4}}(1-\alpha-\beta)K(\alpha,\beta)
[r(m_{Q},s)-5m_{Q}m_{s}]r(m_{Q},s)^{4},\nonumber\\
\rho^{\langle\bar{s}s\rangle}(s)&=&\frac{\langle\bar{s}s\rangle}{3\cdot2^{6}\pi^{4}}\{\int_{\alpha_{min}}^{\alpha_{max}}\frac{d\alpha}{\alpha^{2}}\int_{\beta_{min}}^{1-\alpha}\frac{d\beta}{\beta^{2}}\{[(2-\alpha-\beta)m_{Q}+(1-\alpha-\beta)m_{s}]r(m_{Q},s)\nonumber\\
&&{}-3(\alpha-\alpha^{2}+\beta-\beta^{2})m_{s}m_{Q}^{2}\}r(m_{Q},s)^{2}-m_{s}\int_{\alpha_{min}}^{\alpha_{max}}\frac{d\alpha}{\alpha(1-\alpha)}[m_{Q}^{2}-\alpha(1-\alpha)s]^{3}\},\nonumber\\
\rho^{\langle\bar{s}s\rangle^{2}}(s)&=&\frac{m_{Q}\langle\bar{s}s\rangle^{2}}{3\cdot2^{4}\pi^{2}}\int_{\alpha_{min}}^{\alpha_{max}}d\alpha\{-2m_{Q}[m_{Q}^{2}-\alpha(1-\alpha)s]+m_{s}[m_{Q}^{2}-2\alpha(1-\alpha)s]\},\nonumber\\
\rho^{\langle g\bar{s}\sigma\cdot G s\rangle}(s)&=&\frac{\langle
g\bar{s}\sigma\cdot G
s\rangle}{3\cdot2^{8}\pi^{4}}\{\int_{\alpha_{min}}^{\alpha_{max}}\frac{d\alpha}{\alpha^{2}}\int_{\beta_{min}}^{1-\alpha}\frac{d\beta}{\beta^{2}}r(m_{Q},s)\{-3m_{Q}(\alpha+\beta-4\alpha\beta)r(m_{Q},s)\nonumber\\
&&{}+m_{s}\alpha\beta[12m_{Q}^{2}-7(\alpha+\beta)m_{Q}^{2}-5\alpha\beta s]\}\nonumber\\
&&{}+\int_{\alpha_{min}}^{\alpha_{max}}d\alpha[m_{Q}^{2}-\alpha(1-\alpha)]\{\frac{3m_{Q}}{\alpha(1-\alpha)}[m_{Q}^{2}-\alpha(1-\alpha)s]+2m_{s}[5\alpha(1-\alpha)s-9m_{Q}^{2}]\}\},\nonumber\\
\rho^{\langle g^{2}G^{2}\rangle}(s)&=&-\frac{m_{Q}\langle
g^{2}G^{2}\rangle}{3^{2}\cdot2^{12}\pi^{6}}\int_{\alpha_{min}}^{\alpha_{max}}\frac{d\alpha}{\alpha^{4}}\int_{\beta_{min}}^{1-\alpha}\frac{d\beta}{\beta^{4}}(1-\alpha-\beta)(\alpha^{3}+\beta^{3})K(\alpha,\beta)
r(m_{Q},s)\nonumber\\
&&{}\times[(m_{Q}-3m_{s})r(m_{Q},s)-2m_{s}m_{Q}^{2}(\alpha+\beta)],~\mbox{and}\nonumber\\
\rho^{\langle g^{3}G^{3}\rangle}(s)&=&-\frac{\langle
g^{3}G^{3}\rangle}{3^{2}\cdot2^{14}\pi^{6}}\int_{\alpha_{min}}^{\alpha_{max}}\frac{d\alpha}{\alpha^{4}}\int_{\beta_{min}}^{1-\alpha}\frac{d\beta}{\beta^{4}}(1-\alpha-\beta)K(\alpha,\beta)
\{[(\alpha^{3}+\beta^{3})r(m_{Q},s)+4(\alpha^{4}+\beta^{4})m_{Q}^{2}\nonumber\\
&&{}-2m_{Q}m_{s}(2\alpha^{2}+3\alpha\beta+2\beta^{2})(3\alpha^{2}-4\alpha\beta+3\beta^{2})]r(m_{Q},s)-4m_{s}m_{Q}^{3}(\alpha+\beta)(\alpha^{4}+\beta^{4})\}.\nonumber
\end{eqnarray}
It is defined as $r(m_{Q},s)=(\alpha+\beta)m_{Q}^2-\alpha\beta s$ and $K(\alpha,\beta)=1+\alpha-2\alpha^{2}+\beta+2\alpha\beta-2\beta^{2}$. The integration limits are given
by $\alpha_{min}=(1-\sqrt{1-4m_{Q}^{2}/s})/2$,
$\alpha_{max}=(1+\sqrt{1-4m_{Q}^{2}/s})/2$, and $\beta_{min}=\alpha
m_{Q}^{2}/(s\alpha-m_{Q}^{2})$.
%%%%%%%%%%%%%%%%%%%%%%%%%%%%%%%%%%%%%%%%%%%%%%%%%%%%%%%%%%%%%%%%%%%
\section{Numerical analysis and discussions}\label{sec3}
In this section, the sum rule (\ref{sum rule})
will be numerically simulated.
The input parameters are taken as
$\langle\bar{q}q\rangle=-(0.23\pm0.03)^{3}~\mbox{GeV}^{3}$,
$\langle\bar{s}s\rangle=0.8~\langle\bar{q}q\rangle$, $\langle
g\bar{s}\sigma\cdot G s\rangle=m_{0}^{2}~\langle\bar{s}s\rangle$,
$m_{0}^{2}=0.8~\mbox{GeV}^{2}$, $\langle
g^{2}G^{2}\rangle=0.88~\mbox{GeV}^{4}$, and $\langle
g^{3}G^{3}\rangle=0.045~\mbox{GeV}^{6}$ \cite{overview2,Y4660-QCDSR,parameters}.
For the quark masses, we employ the same values as Ref. \cite{Narison} and references therein, which spanned by the
running $\overline{MS}$ mass and the on-shell mass from QCD sum rule, with
$m_{c}=1.26\sim1.47~\mbox{GeV}$, $m_{b}=4.22\sim4.72~\mbox{GeV}$,
as well as $m_{s}=114.5\pm20.8~\mbox{MeV}$.
Complying with the standard criterion
of sum rule analysis, the threshold $s_{0}$ and Borel
parameter $M^{2}$ are varied to find the stability window.
It is well known that the fundamental assumption
of the QCD sum rule is the principle of duality:
it is assumed that there is an interval over which a
hadron may be equivalently described at both the
quark level and the hadron level. Therefore, the correlation function
is evaluated in two different ways: at the quark level
in terms of quark and gluon fields and at the hadronic
level.
If both sides of the sum rule were calculated to
arbitrarily high accuracy, the matching of them would be independent of $M^{2}$.
Practically, however, both sides are represented
imperfectly. On one hand, there are
approximations in the OPE of the correlation functions and, on the other
hand, there is a very complicated and largely unknown structure of the hadronic
dispersion integrals in the phenomenological side. Thus,
the extracted result is not completely independent of $M^{2}$.
The hope is that there exists a range of $M^{2}$,
in which the two sides have a good
overlap and information on the resonance can be
extracted. In practice, one can analyse the OPE convergence and the pole contribution to determine the
allowed Borel window of $M^{2}$: the lower limit
constraint for $M^{2}$ is obtained
by restricting that the
perturbative contribution should be larger than the condensate contributions;
the upper limit constraint is gained by the consideration
that the pole contribution should be larger
than QCD continuum contribution.
Meanwhile,
the threshold parameter
$\sqrt{s_{0}}$ characterizes the beginning
of the continuum state.
Thereby, it is not
arbitrary but correlated to the energy of the next excited state with the same quantum number as the studied state.

At first, we keep the values of the quark masses and condensates fixed at the central values.
The comparison
between pole and continuum contributions from sum rule (\ref{sumrule}) for $Y_{[cs]}$
for $\sqrt{s_{0}}=5.2~\mbox{GeV}$ is shown in the left part of FIG. 1, and its OPE convergence by comparing the
perturbative, quark condensate, four-quark condensate, mixed condensate, two-gluon condensate, and three-gluon
condensate contributions is shown in the right one. Numerically, the ratio of
perturbative contribution to the total OPE contribution at $M^{2}=2.5~\mbox{GeV}^{2}$ is nearly $60\%$, which is increasing
with the $M^{2}$ to insure that perturbative contribution can dominate in the total OPE contribution
when $M^{2}\geq2.5~\mbox{GeV}^{2}$. On the other side, the relative pole contribution
is approximate to $52\%$ at $M^{2}=3.2~\mbox{GeV}^{2}$ and descending along with the $M^{2}$
to guarantee the pole contribution can dominate in the total contribution while $M^{2}\leq3.2~\mbox{GeV}^{2}$.
Thus, the region of $M^{2}$ for $Y_{[cs]}$ is taken as
$M^{2}=2.5\sim3.6~\mbox{GeV}^{2}$ for $\sqrt{s_0}=5.2~\mbox{GeV}$.
Similarly,
the proper range of $M^{2}$ is gained as $2.5\sim3.0~\mbox{GeV}^{2}$ for $\sqrt{s_0}=5.0~\mbox{GeV}$, and
the range of $M^{2}$ is $2.5\sim3.6~\mbox{GeV}^{2}$ for $\sqrt{s_0}=5.4~\mbox{GeV}$.
We see also that for $\sqrt{s_0}=4.9~\mbox{GeV}$,
the corresponding Borel parameter range is $M^{2}=2.5\sim2.7~\mbox{GeV}^{2}$,
which is very narrow as a working window. It is the main reason that $\sqrt{s_0}\leq4.9~\mbox{GeV}$ is not chosen here.
In order to evaluate the uncertainty of results more conservatively \cite{FESR},
we enlarge the variation of threshold parameter $\sqrt{s_{0}}$ for $Y_{cs}$ from $5.0\sim5.4~\mbox{GeV}$ to $5.0\sim5.7~\mbox{GeV}$ 
and we find the range of $M^{2}$ is $2.5\sim3.8~\mbox{GeV}^{2}$ for $\sqrt{s_0}=5.7~\mbox{GeV}$.
In the chosen region,
the mass result is not completely independent of $M^{2}$ since both sides of the sum rule are not calculated to
arbitrarily high accuracy but have included some approximations, and that is
just the reason by which the accuracy of QCD sum rule method
is limited. Whereas,
it is expected that the two sides have a good
overlap and information on the resonance can be safely
extracted in the chosen range of $M^{2}$.
The corresponding Borel curve to determine the mass of $Y_{[cs]}$ is exhibited in the left part of FIG. 3.
We compute the average mass value of these working windows as
$4.69\pm0.29~\mbox{GeV}$ (the numerical error reflects the uncertainty due
to variation of $s_{0}$ and
$M^{2}$).
Up to now, we have kept the values of the quark masses and condensates at the central values.
At last, we vary the quark masses as well as condensates and arrive at
$4.69\pm0.29\pm0.07~\mbox{GeV}$ (the first error reflects the uncertainty due
to variation of $s_{0}$ and
$M^{2}$, and the second error resulted from the variation of QCD parameters) or $4.69\pm0.36~\mbox{GeV}$ in a concise form.

For $Y_{[bs]}$,
the comparison
between pole and continuum contributions from sum rule (\ref{sumrule}) for $\sqrt{s_0}=11.8~\mbox{GeV}$
is shown in the left part of FIG. 2, and its OPE convergence by comparing different OPE
contributions is shown in the right one. In detail, the
perturbative contribution versus the total OPE contribution at $M^{2}=7.5~\mbox{GeV}^{2}$ is nearly $62\%$,
and the relative pole contribution
is approximate to $50\%$ at $M^{2}=9.0~\mbox{GeV}^{2}$.
Thus, the region of $M^{2}$ is
taken as $M^{2}=7.5\sim9.0~\mbox{GeV}^{2}$ for $\sqrt{s_0}=11.8~\mbox{GeV}$.
With the similar analysis,
for $\sqrt{s_0}=11.6~\mbox{GeV}$, the range is $M^{2}=7.5\sim8.3~\mbox{GeV}^{2}$;
for $\sqrt{s_0}=12.0~\mbox{GeV}$, the range is $M^{2}=7.5\sim9.5~\mbox{GeV}^{2}$.
To evaluate the uncertainty of results more conservatively,
we enlarge
the variation of $\sqrt{s_{0}}$ from $11.6\sim12.0~\mbox{GeV}$ to $11.6\sim12.3~\mbox{GeV}$.
For $\sqrt{s_0}=12.3~\mbox{GeV}$, the range of $M^{2}$ is $7.5\sim10.3~\mbox{GeV}^{2}$.
The
dependence on $M^2$ for the mass of $Y_{[bs]}$ from sum rule (\ref{sum rule}) is
shown in the right part of FIG. 3. For $Y_{bs}$,
We arrive at
$11.19\pm0.28~\mbox{GeV}$ (not including the variation of QCD parameters).
Finally, we vary the quark masses as well as condensates and arrive at
$11.19\pm0.28\pm0.21~\mbox{GeV}$ (the former error reflects the uncertainty due
to variation of $s_{0}$ and
$M^{2}$, and the latter error resulted from the variation of QCD parameters) or $11.19\pm0.49~\mbox{GeV}$
in a concise form.

With regard to the numerical results, some more discussions are given below. Numerically,
the result $4.69\pm0.36~\mbox{GeV}$ for $Y_{[cs]}$ is in good agreement with the experimental value
$4664\pm11\pm5~\mbox{MeV}$ for $Y(4660)$. However, its value is
a bit higher than $Y(4260)$'s mass even considering the uncertainty,
which supports the $P$-wave $[cs][\bar{c}\bar{s}]$ structure for $Y(4660)$ while disfavors the
explanation of $Y(4260)$ as the $P$-wave
$[cs][\bar{c}\bar{s}]$ state. Note that
some authors also assume that $Y(4260)$ could be a $P$-wave $[cq][\bar{c}\bar{q}]$ state \cite{Y4660-Ebert}.
In fact, we have calculated the mass
of the $P$-wave $[cq][\bar{c}\bar{q}]$ to be $4.32\pm0.20~\mbox{GeV}$ \cite{jrz},
which is compatible with the experimental data
of $Y(4360)$ and could
support $Y(4360)$'s $P$-wave $[cq][\bar{c}\bar{q}]$ structure.
Barely from the value $4.32\pm0.20~\mbox{GeV}$, one could not completely exclude the possibility of $Y(4260)$ as
a $P$-wave $[cq][\bar{c}\bar{q}]$ state
since it is still in accord with the mass of $Y(4260)$ in view of the
uncertainty. Concerning the real nature of
$Y(4260)$, some further theoretical study and experimental verification
are undoubtedly needed.

\begin{figure}
\centerline{\epsfysize=5.8truecm\epsfbox{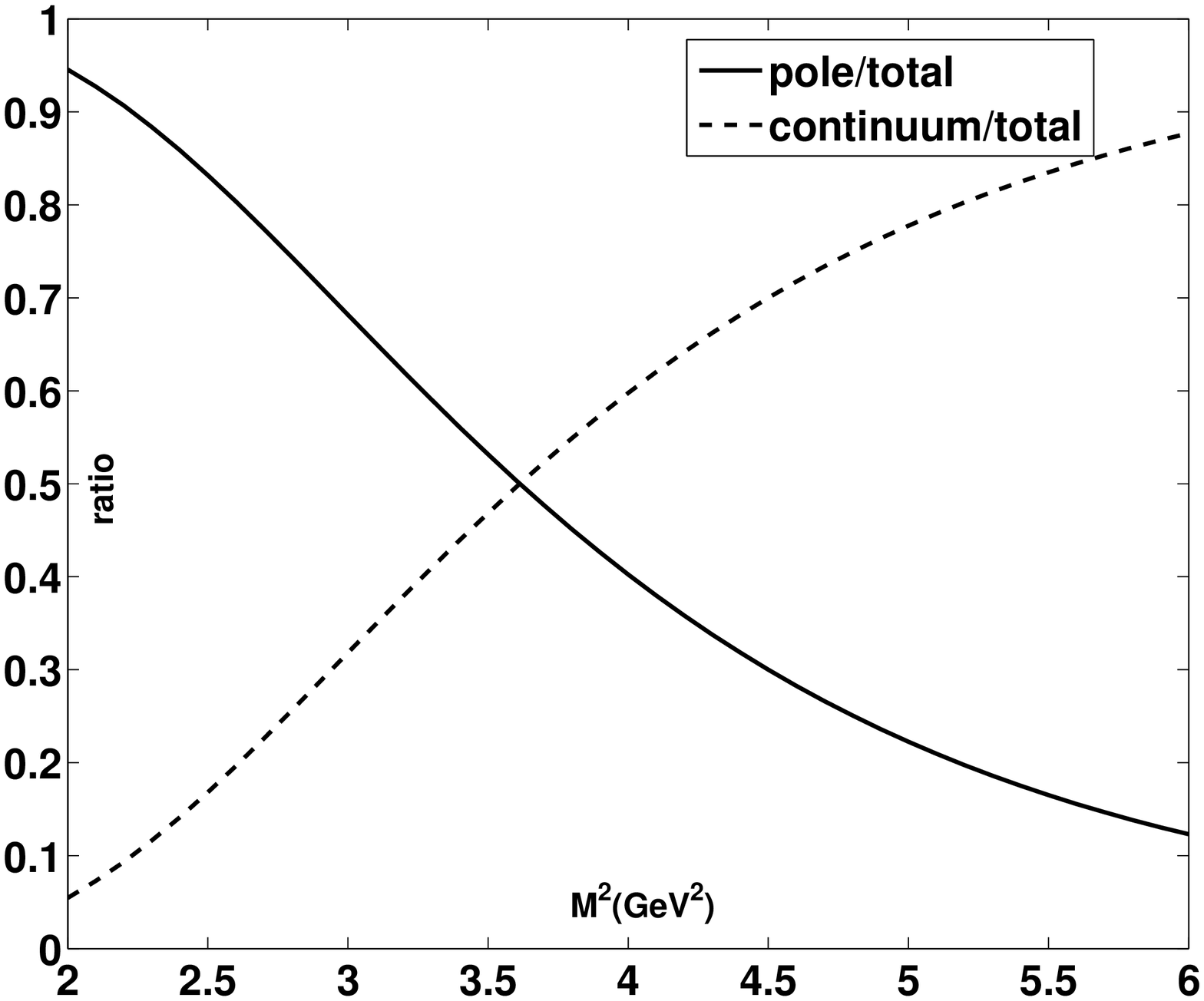}\epsfysize=5.8truecm\epsfbox{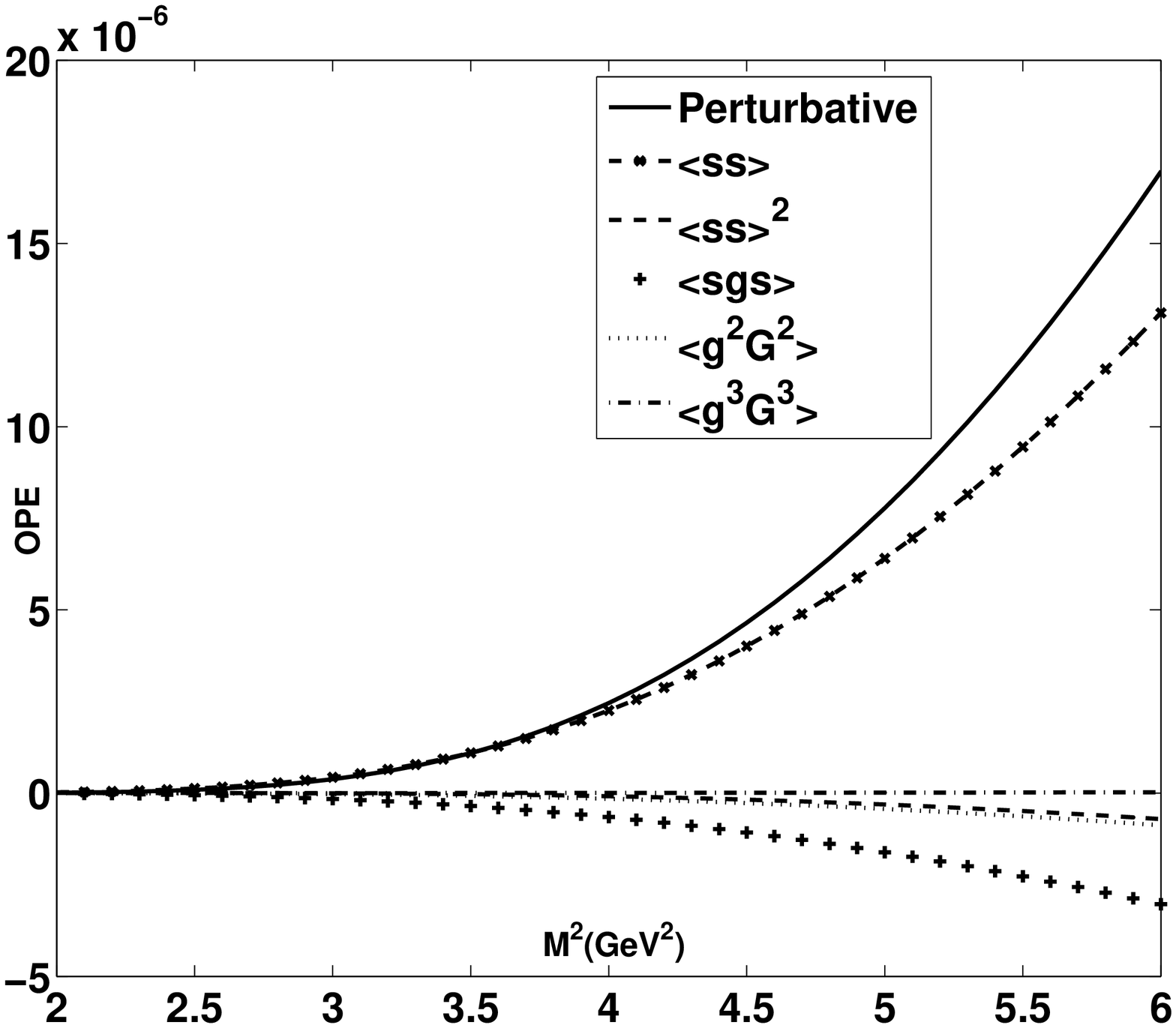}}
\caption{In the left part, the solid line shows the relative pole contribution
(the pole contribution divided by the total, pole plus continuum
contribution) and the dashed line shows the relative continuum
contribution from sum rule (\ref{sumrule}) for $\sqrt{s_{0}}=5.2~\mbox{GeV}$ for
$Y_{[cs]}$. The OPE convergence is shown by comparing the
perturbative, quark condensate, four-quark condensate, mixed condensate, two-gluon condensate and three-gluon
condensate contributions from sum rule (\ref{sumrule}) for $\sqrt{s_{0}}=5.2~\mbox{GeV}$ for
$Y_{[cs]}$ in the right one. }
\end{figure}

\begin{figure}
\centerline{\epsfysize=5.8truecm\epsfbox{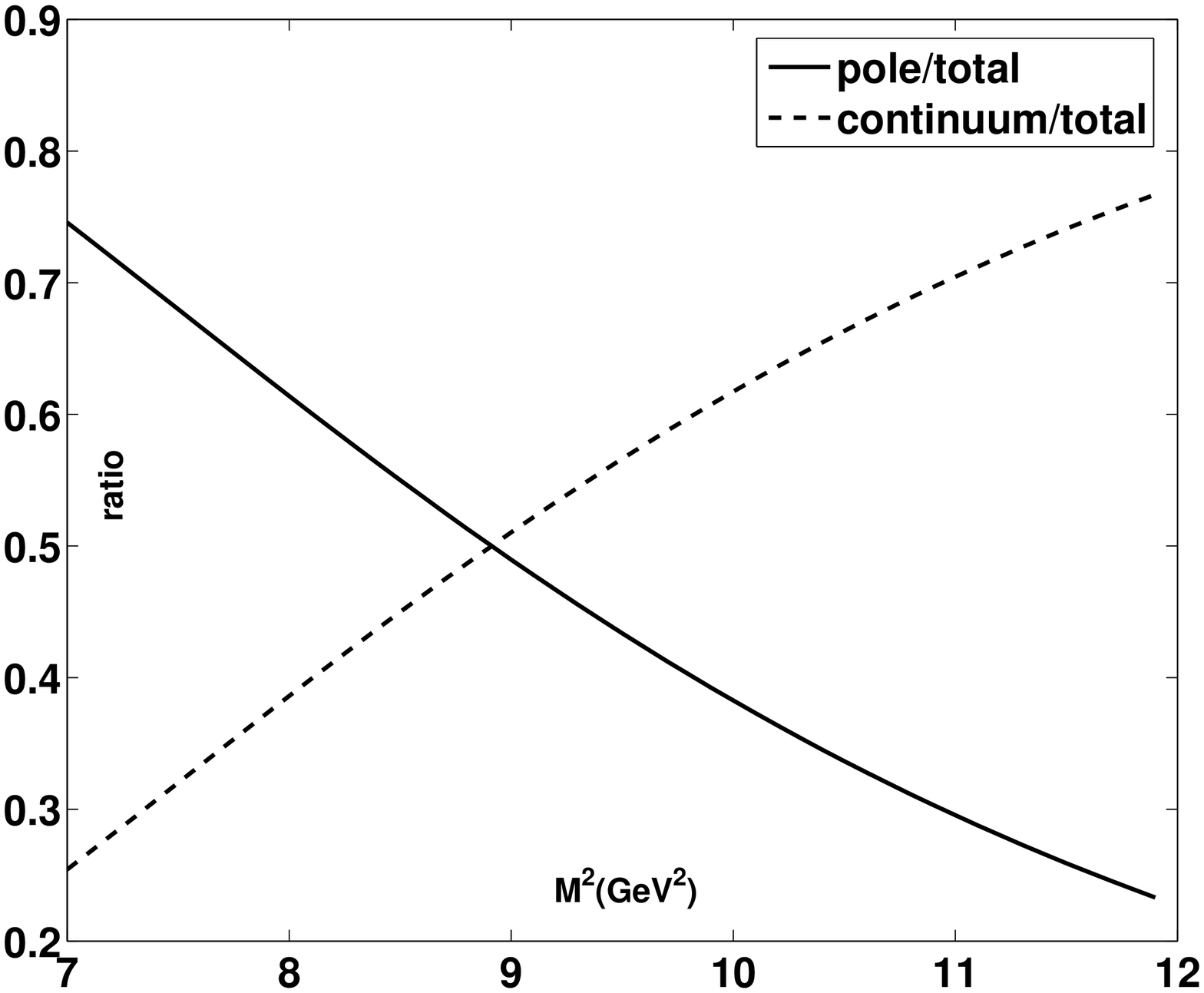}\epsfysize=5.8truecm\epsfbox{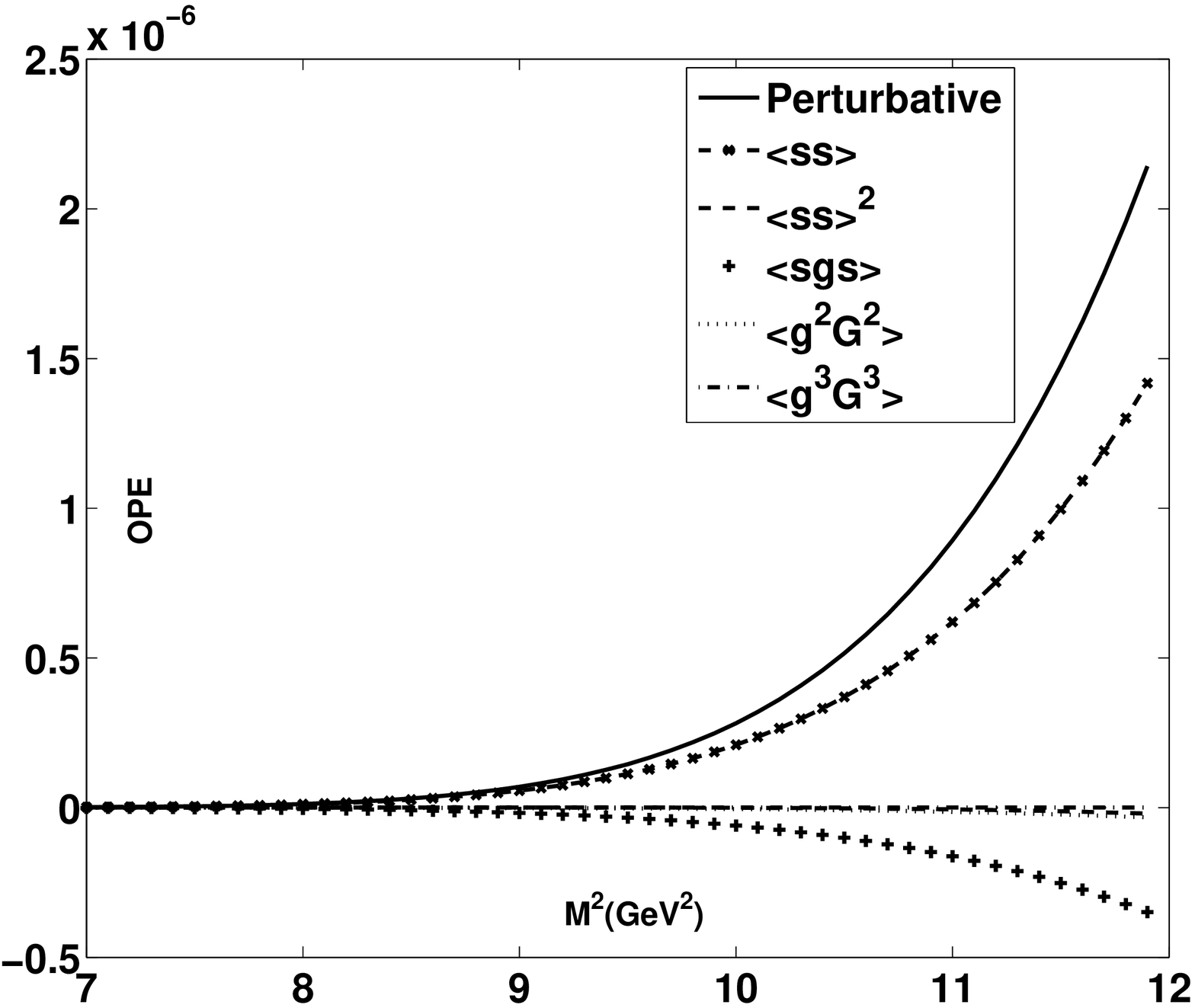}}
\caption{In the left part, the solid line shows the relative pole contribution
(the pole contribution divided by the total, pole plus continuum
contribution) and the dashed line shows the relative continuum
contribution from sum rule (\ref{sumrule}) for $\sqrt{s_{0}}=11.8~\mbox{GeV}$ for
$Y_{[bs]}$. The OPE convergence is shown by comparing the
perturbative, quark condensate, four-quark condensate, mixed condensate, two-gluon condensate and three-gluon
condensate contributions from sum rule (\ref{sumrule}) for $\sqrt{s_{0}}=11.8~\mbox{GeV}$ for
$Y_{[bs]}$ in the right one.}
\end{figure}

\begin{figure}
\centerline{\epsfysize=5.8truecm
\epsfbox{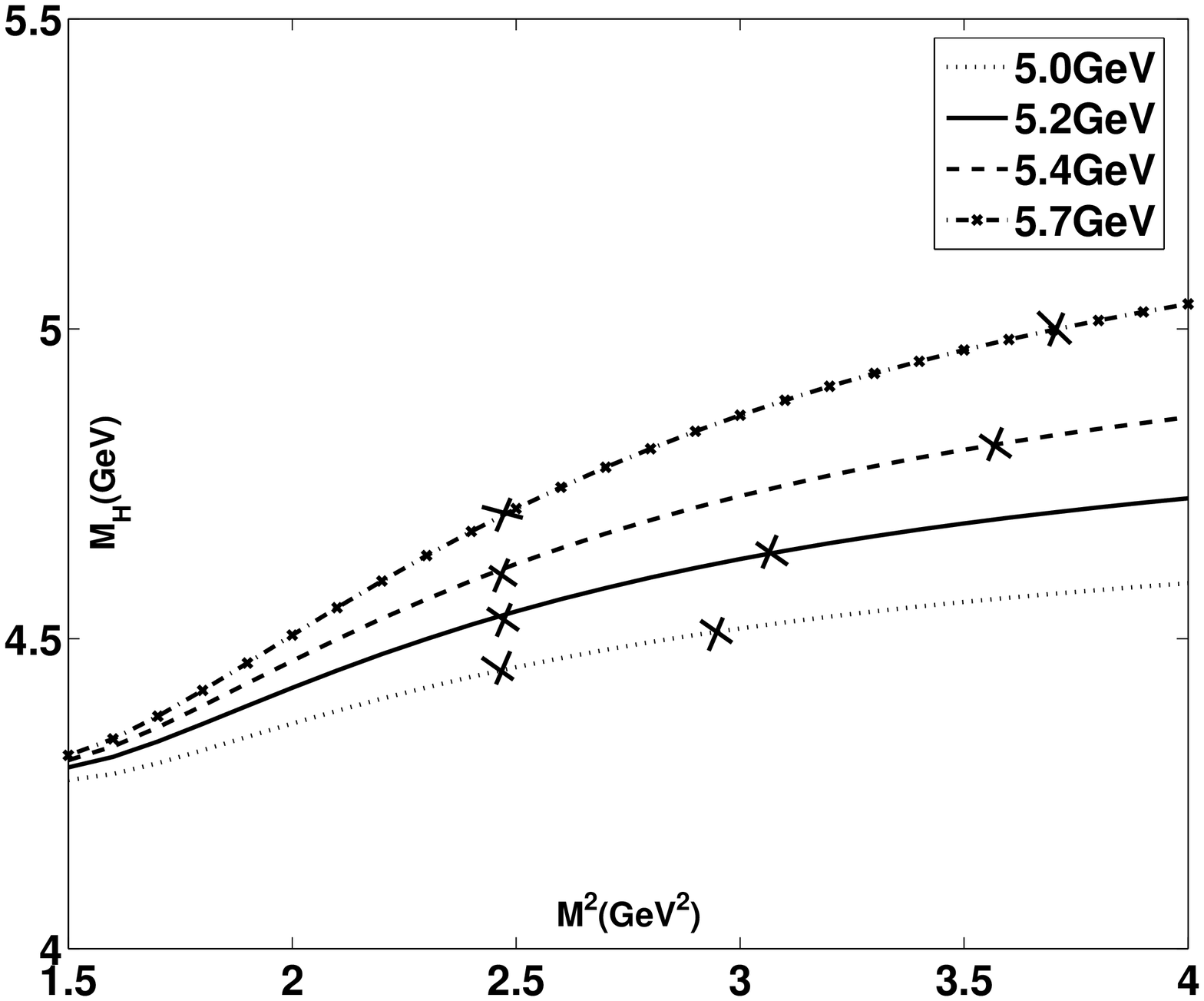}\epsfysize=5.8truecm
\epsfbox{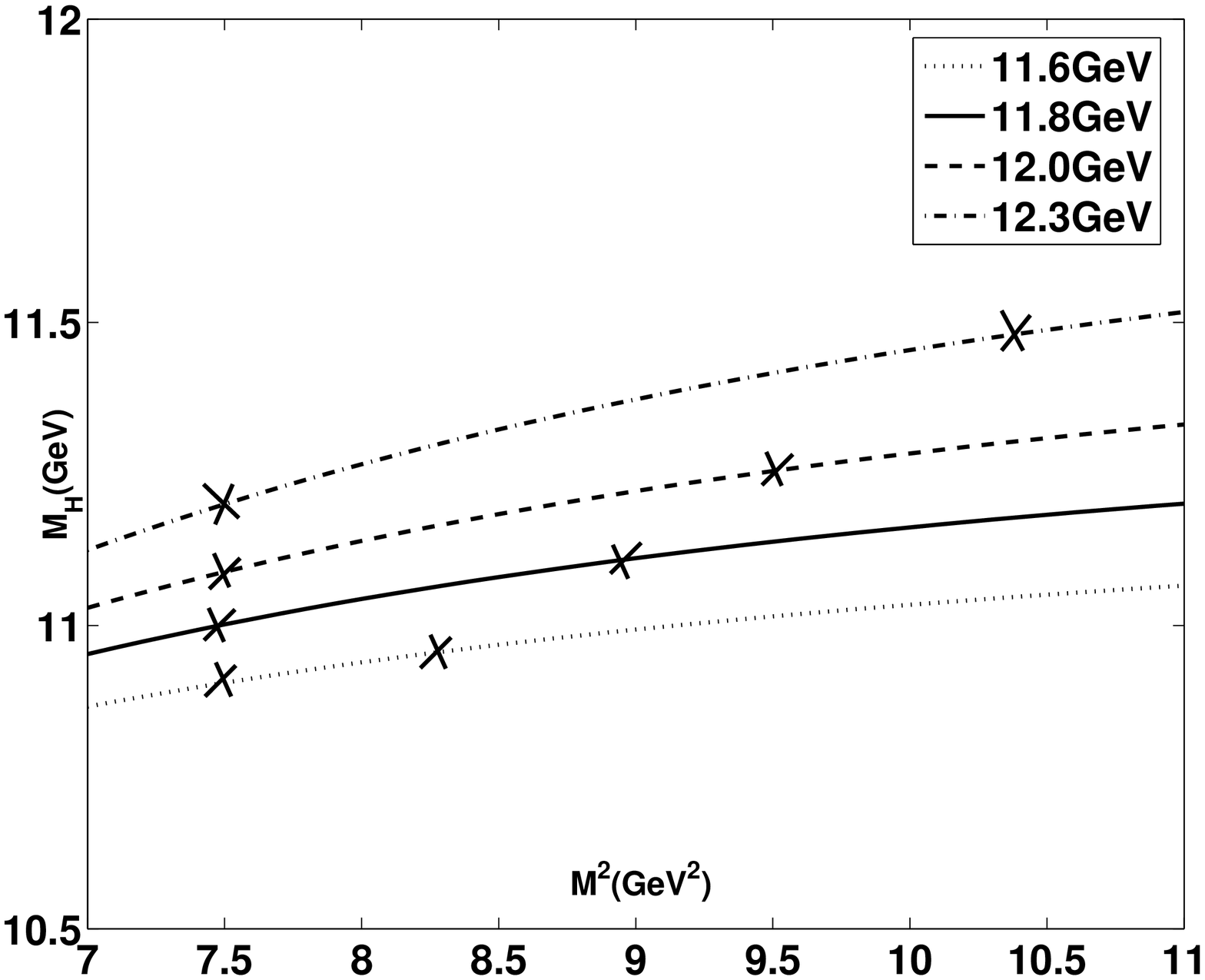}}\caption{In the left part, the
dependence on $M^2$ for the mass of $Y_{[cs]}$ from sum rule (\ref{sum rule}) is shown. The continuum
thresholds are taken as $\sqrt{s_0}=5.0\sim5.7~\mbox{GeV}$. For
$\sqrt{s_0}=5.0~\mbox{GeV}$, the range of $M^{2}$ is $2.5\sim3.0~\mbox{GeV}^{2}$;
for $\sqrt{s_0}=5.2~\mbox{GeV}$, the range of $M^{2}$ is $2.5\sim3.2~\mbox{GeV}^{2}$;
for $\sqrt{s_0}=5.4~\mbox{GeV}$, the range of $M^{2}$ is $2.5\sim3.6~\mbox{GeV}^{2}$;
for $\sqrt{s_0}=5.7~\mbox{GeV}$, the range of $M^{2}$ is $2.5\sim3.8~\mbox{GeV}^{2}$. The
dependence on $M^2$ for the mass of $Y_{[bs]}$ from sum rule (\ref{sum rule}) is shown in the right one. The continuum
thresholds are taken as $\sqrt{s_0}=11.6\sim12.3~\mbox{GeV}$. For
$\sqrt{s_0}=11.6~\mbox{GeV}$, the ranges of $M^{2}$ is $7.5\sim8.3~\mbox{GeV}^{2}$;
for $\sqrt{s_0}=11.8~\mbox{GeV}$, the range of $M^{2}$ is $7.5\sim9.0~\mbox{GeV}^{2}$;
for $\sqrt{s_0}=12.0~\mbox{GeV}$, the range of $M^{2}$ is $7.5\sim9.5~\mbox{GeV}^{2}$;
for $\sqrt{s_0}=12.3~\mbox{GeV}$, the range of $M^{2}$ is $7.5\sim10.3~\mbox{GeV}^{2}$.}
\end{figure}

%%%%%%%%%%%%%%%%%%%%%%%%%%%%%%%%%%%%%%%%%%%%%%%%%%%%%%%%%%%%%%%%%%%
\section{Summary}\label{sec4}
The QCD sum rule method has been employed to compute the
mass of $P$-wave $[cs][\bar{c}\bar{s}]$ tetraquark state $Y_{[cs]}$, including
contributions of operators up to dimension six in the OPE.
The final result $4.69\pm0.36~\mbox{GeV}$ ($4.69\pm0.29\pm0.07~\mbox{GeV}$, where the first error reflects the uncertainty due
to variation of $s_{0}$ and
$M^{2}$, and the second error resulted from the variation of QCD parameters) for
$Y_{[cs]}$ is well compatible with the experimental data
of $Y(4660)$, which favors the $P$-wave tetraquark configuration for $Y(4660)$.
Meanwhile, the result is higher than $Y(4260)$'s mass, which
is not consistent with assumption of $Y(4260)$ as the $P$-wave
$[cs][\bar{c}\bar{s}]$ state.
As a byproduct, the mass for the bottom counterpart $Y_{[bs]}$
has also been predicted, which is $11.19\pm0.49~\mbox{GeV}$ ($11.19\pm0.28\pm0.21~\mbox{GeV}$, where the former error reflects the uncertainty due
to variation of $s_{0}$ and
$M^{2}$, and the latter error resulted from the variation of QCD parameters) and expecting further experimental
identification.

%%%%%%%%%%%%%%%%%%%%%%%%%%%%%%%%%%%%%%
\begin{acknowledgments}
This work was supported by the National Natural Science
Foundation of China under Contract No.10975184.
\end{acknowledgments}
%%%%%%%%%%%%%%%%%%%%%%%%%%%%%%%%%%%%%%%%%%%%%%%%%%%%

\end{document}